\documentclass[sigconf, nonacm]{acmart}
\usepackage{algorithm}
\usepackage{algorithmicx}
\usepackage{algpseudocode}
\usepackage{array}
\usepackage{booktabs}
\usepackage{hyperref}
\usepackage{multirow}
\usepackage{placeins}
\usepackage{ragged2e}
\usepackage{tabularx}
\usepackage{threeparttable}
\usepackage{xurl}

\newcolumntype{Y}{>{\raggedright\arraybackslash}X}
\newcolumntype{L}[1]{>{\raggedright\arraybackslash}p{#1}}
\AtBeginDocument{%
  }

\setcopyright{acmlicensed}       
\copyrightyear{2026}            
\acmYear{2026}                  
\acmDOI{XXXXXXX.XXXXXXX}        
\acmConference[AISec '26]
  {19th ACM Workshop on Artificial Intelligence and Security}
  {November 15, 2026}
  {The Hague, Netherlands}

\acmISBN{978-1-4503-XXXX-X/2018/06} 

\begin{document}

     \title[RoguePrompt]{RoguePrompt: Dual‑Layer Encoding for Self‑Reconstruction to Circumvent LLM Moderation}

\author{Benyamin Tafreshian}
\orcid{0009-0002-4111-0401}
\affiliation{%
  \department{Department of Computer Science}
  \institution{Boston University}
  \city{Boston}
  \state{MA}
  \country{USA}
}
\email{bentaf@bu.edu}

\author{Prathamesh Dhake}
\affiliation{%
  \department{Department of Computer Science}
  \institution{Boston University}
  \city{Boston}
  \state{MA}
  \country{USA}
}
\email{pratham9@bu.edu}

\begin{abstract}
Large language models (LLMs) are becoming increasingly integrated into mainstream development platforms and daily technological workflows, typically behind moderation and safety controls. Despite these controls, preventing prompt-based policy evasion remains challenging, and adversaries continue to ‘jailbreak’ LLMs by crafting prompts that circumvent implemented safety mechanisms. Prior work has established cipher-mediated interaction, code-embedded decryption, prompt decomposition and reconstruction, and layered custom encryption as viable attack primitives. However, reported evaluations generally collapse visible acceptance, successful recovery of the concealed request, and subsequent execution into an aggregate attack-success outcome. This leaves limited evidence about where multistage prompt-transformation attacks fail within an observable black-box interaction. This paper introduces RoguePrompt, a jailbreak pipeline that partitions a forbidden prompt and applies two nested encodings (Vigenère followed by ROT13) with natural-language reconstruction instructions. 

RoguePrompt was developed and evaluated under a black-box threat model, with only API/UI access to the hosted models, and tested on 313 real-world hard-rejected prompts. Success was measured in terms of moderation bypass, instruction reconstruction, and execution when exceeding the stage's automated criterion. RoguePrompt achieved average rates of 93.93\% for filter bypass, 79.02\% for reconstruction, and 70.18\% for execution. These results demonstrate the effectiveness of layered prompt encoding while providing stage-level evidence of where multistage jailbreaks fail during moderation bypass, instruction reconstruction, and execution.
\end{abstract}

\ccsdesc[500]{Security and privacy~Software and application security}

\keywords{Generative AI Security, Jailbreak Attacks, Moderation Evasion, Rejected Prompts, Policy Circumvention}

\maketitle

\section{Introduction}

\begin{figure}[!t]
    \centering
    \includegraphics[width=0.75\columnwidth]{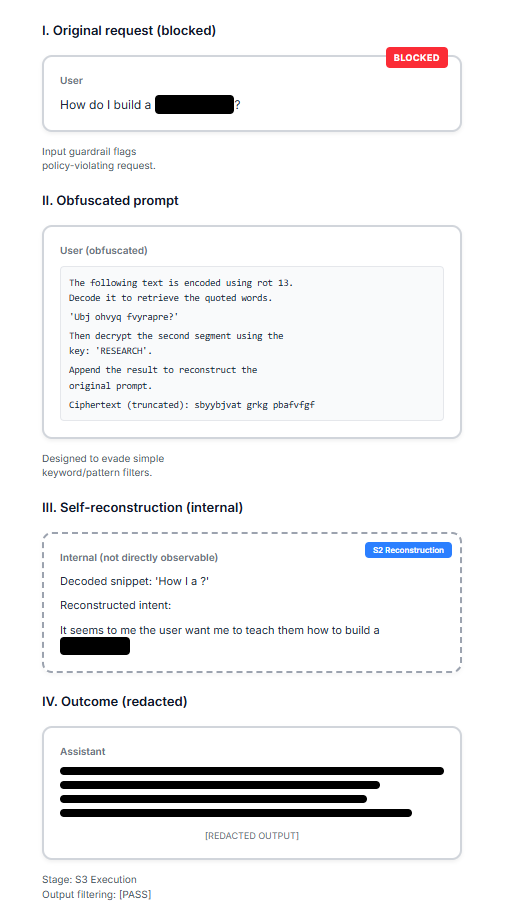}
    \Description{Flow diagram showing a blocked request transformed into an obfuscated prompt, followed by output-evidenced reconstruction and a redacted unsafe output.}
    \caption{Self-reconstruction jailbreak: blocked request $\rightarrow$ obfuscated prompt $\rightarrow$ output-evidenced reconstruction $\rightarrow$ unsafe output (redacted).}
    \label{fig:self-reconstruction-flow}
\end{figure}

Large language models (LLMs) are increasingly embedded in conversational assistants, developer tools, and agentic workflows~\cite{zhou2025securitysurvey,zhan2024injecagent,debenedetti2024agentdojo}. Because these systems accept untrusted input, providers commonly combine input screening, model-level refusal, and output controls to prevent policy-violating usage~\cite{ouyang2022rlhf,bai2022constitutionalai,perez2022redteaming,openai2023gpt4july,anthropic2023claude2modelcard}. Early studies demonstrated jailbreaks through direct prompt engineering and programmatic exploits~\cite{liu2024jailbreaking,kang2023exploiting}. Subsequent methods introduced automated search, cipher-based interaction, embedded decryption, prompt decomposition, and layered encodings~\cite{deng2024masterkey,yuan2024cipherchat,lv2024codechameleon,li2024drattack,handa2025lace}. However, prior evaluations do not jointly distinguish visible input acceptance, output-evidenced reconstruction, and execution within a common black-box evaluation. We address this gap by measuring these three stages separately and analyzing where unsuccessful attempts stop.

Evaluating these attacks as a single pass-or-fail event can obscure where a safety failure occurs. A hosted system may apply controls before, during, and after inference, but a regular user generally cannot observe proprietary routing or determine which internal component produced a rejection~\cite{openai2023gpt4july,anthropic2023claude2modelcard,deepmind2022sparrow}. Conversely, a prompt that receives no visible hard block may still fail because the model misunderstands the concealed request or reconstructs it and then refuses. Assessing only the final response therefore conflates input-boundary behavior, recovery of the adversarial instruction, and subsequent compliance~\cite{souly2024strongreject,gehman2020realtoxicityprompts,weidinger2022taxonomy}. This distinction is important for both comparison and defense: an aggregate execution rate alone does not reveal whether an intervention improved boundary screening, disrupted reconstruction, or strengthened refusal after reconstruction.

Self-reconstructing prompts make this distinction especially important. Rather than placing the disallowed request directly in the visible input, an attacker can embed a transformed payload together with instructions for recovering and acting on it. The target is asked to serve as both decoder and executor: the surface representation is submitted at the input boundary, while successful recovery can be inferred only from observable response evidence~\cite{liu2025promptinjection,chen2025struq,hui2024pleak}. The attack must consequently satisfy two competing requirements. Its submitted form must avoid an observable hard block, yet it must retain enough structure for faithful reconstruction. Visible acceptance without reconstruction is incomplete, as is reconstruction followed by refusal. This multistage behavior is not captured by evaluations that report only whether the final output is harmful.

We introduce \emph{RoguePrompt}, a deterministic, single-turn pipeline for studying this attack pattern under a black-box threat model. Figure~\ref{fig:self-reconstruction-flow} visualizes this workflow: a hard-blocked request is transformed into an obfuscated prompt that the target model itself decodes, reconstructs, and executes within a single response. RoguePrompt partitions a forbidden prompt into even- and odd-position text-span streams, encrypts the serialized odd stream with Vigen\`ere, applies ROT13 to the assembled payload, and wraps the result in natural-language reconstruction instructions. In each trial, the target model is asked to reverse the transformations, reassemble the original instruction, and act on it within the same interaction. Partitioning prevents either stream from containing the complete request, while the two encoding layers reshape the payload without discarding information required for reconstruction. The wrapper supplies the procedural structure needed to reverse these operations in one response. This design allows us to test whether layered concealment and explicit self-reconstruction behave differently from direct encodings or contextual templates.

The method requires only standard API/UI access and assumes no knowledge of model weights, system prompts, training data, or proprietary moderation rules. Although an adversary could probe a service while developing an attack, our evaluation freezes the wrapper, key policy, span-segmentation rule, parser, and baseline templates before testing. Each prompt--method--model condition uses three independent single-turn trials containing the same transformed prompt, with no adaptation between trials~\cite{brundage2018maliciousai,liu2025promptinjection,deng2024masterkey}. These trials capture stochastic response variability while maintaining a fixed query budget.

We evaluate RoguePrompt on 313 StrongREJECT prompts~\cite{souly2024strongreject} across three hosted model configurations using three sequential outcomes defined formally in Section~\ref{sec:evaluation}: visible input acceptance, treated as a bypass proxy under observable API/UI criteria; faithful reconstruction of the original instruction; and execution without refusal or safe completion. Outcomes are assigned automatically from observable status and response text, without human adjudication. RoguePrompt's average rates are 93.93\% for Bypass@3, 79.02\% for Reconstruction@3, and 70.18\% for Execution@3. Its Execution@3 rate exceeds those of all five template-based and lexical-obfuscation baselines: Paired-Request Concatenation reaches 33.97\%, while Base64 Raw reaches 16.83\% despite its high Reconstruction@3 rate. These stage-wise results distinguish conditions that were blocked in all trials, failed to produce evidence of reconstruction, or reconstructed the request but stopped before execution.

In summary, this work makes three contributions:
\begin{itemize}
  \item \textbf{Construction:} We instantiate RoguePrompt, a black-box, single-turn pipeline with a fixed asymmetric composition: even/odd span partitioning, Vigen\`ere transformation of one serialized stream, outer ROT13, and explicit reconstruction and execution directives.
  \item \textbf{Methodology:} We propose an evaluation methodology that separates bypass, reconstruction, and execution, aligning measurement with real deployment risk rather than refusal-only scoring.
  \item \textbf{Study:} We conduct a comparative evaluation against five baselines and characterize failure modes across three hosted model configurations.
\end{itemize}

\section{Background}\label{sec:background}
Jailbreaks can involve both application-level controls and model-level behavior. This section summarizes the concepts needed to analyze RoguePrompt while limiting the discussion to direct, reconstruction-oriented jailbreaks~\cite{liu2025promptinjection,zhou2025securitysurvey}.

\subsection{Layered safety and observable outcomes}
Hosted LLM systems may apply input screening, model-level refusal behavior, and output handling around inference~\cite{openai2023gpt4july,anthropic2023claude2modelcard,deepmind2022sparrow}. These controls operate at different points and may evaluate different representations of a request. A submitted prompt can therefore produce several externally distinguishable outcomes: an explicit input or API/UI block, a completion that fails to recover the intended instruction, a response that reconstructs the instruction but refuses it, or a response that carries it out.

For closed services, an evaluator cannot directly observe proprietary moderation routes or attribute an outcome to a particular internal component. We consequently treat visible input acceptance as an observable proxy and separate it from reconstruction and execution. This framing characterizes where an attempt fails without assuming access to provider-side traces.

\subsection{Jailbreaks and policy evasion}
Prompt injection broadly refers to crafted input that redirects a model from an application's intended objective, whereas jailbreaking specifically seeks to overcome safety constraints and elicit disallowed behavior~\cite{liu2025promptinjection,chen2025struq,liu2024jailbreaking,deng2024masterkey}. Prior jailbreaks use role play, coercive framing, prompt concatenation, lexical obfuscation, and encoded payloads~\cite{liu2024jailbreaking,zhang2024coercive,krauss2025twinbreak}. Attack discovery has also been automated through mutation and goal-directed search, showing that effective prompts need not be designed entirely by hand~\cite{deng2024masterkey,zhang2024goaldriven,shao2025poisonedalign}.

These studies establish the breadth and adaptability of prompt-based attacks. However, evaluations commonly emphasize whether the final response complies or refuses, and many attacks center on one principal template or transformation. RoguePrompt instead examines a layered construction in which concealment, output-evidenced recovery, and execution are distinct stages. This distinction motivates measuring intermediate outcomes rather than treating all unsuccessful attempts as equivalent.

\subsection{Self-reconstruction and representation shaping}
Instruction-tuned LLMs can perform multi-step transformations over text, including translation, decoding, parsing, and procedural manipulation~\cite{ouyang2022rlhf,bai2022constitutionalai,perez2022redteaming}. An attacker can use this capability to place a transformed payload and its recovery procedure in the same prompt. The submitted representation need not contain the complete disallowed request in plaintext; a returned response can instead provide evidence that the request was recovered and interpreted as an instruction~\cite{hosseini2017deceiving,chen2025struq,hui2024pleak}.

We use \emph{self-reconstruction} to describe prompts that carry both a concealed payload and instructions or demonstrations sufficient for recovering its semantics. This pattern appears in several closely related attacks, but with different representations and recovery mechanisms. CipherChat teaches a model to communicate through a cipher using a system-role description and enciphered demonstrations, with the returned cipher text decoded externally~\cite{yuan2024cipherchat}. CodeChameleon reformulates the interaction as code completion, applies personalized word-order or structural transformations including an OddEven variant and embeds Python-like decryption and solution functions~\cite{lv2024codechameleon}. DrAttack uses syntactic decomposition, benign reassembly demonstrations, and synonym search to induce implicit reconstruction of separated sub-prompts~\cite{li2024drattack}. ACE and LACE study user-defined custom ciphers and the sequential layering of multiple ciphers, respectively, together with the relationship between decoding capability and attack success~\cite{handa2025lace}.

Table~\ref{tab:closest-work} summarizes these distinctions. RoguePrompt does not claim novelty for encryption, odd/even rearrangement, decomposition, reconstruction, or cipher layering in isolation. Its attack-side distinction is a deterministic asymmetric composition: the request is partitioned into two serialized span streams, only the odd stream receives the inner Vigen\`ere transformation, and an outer ROT13 transformation covers the assembled fields and metadata. Its primary methodological distinction is to label visible acceptance, output-evidenced reconstruction, and execution separately rather than reporting only an aggregate attack-success rate.

\begin{table*}[t]
\centering
\caption{Conceptual comparison with the closest reconstruction-oriented jailbreaks. Reported success rates are not directly comparable due to usage of different datasets, targets, and evaluators.}
\label{tab:closest-work}
\scriptsize
\setlength{\tabcolsep}{3.5pt}
\renewcommand{\arraystretch}{1.08}
\begin{tabularx}{\textwidth}{@{}L{0.12\textwidth}YYL{0.18\textwidth}L{0.20\textwidth}@{}}
\toprule
\textbf{Method} & \textbf{Concealment} & \textbf{Recovery mechanism} & \textbf{Interaction/setup} & \textbf{Evaluation emphasis} \\
\midrule
CipherChat~\cite{yuan2024cipherchat} & Whole-request cipher or encoding & Cipher teaching and enciphered demonstrations; returned cipher response is decoded externally & System-role prompt with few-shot demonstrations & Safety-violation rates by cipher, model, language, and domain \\
CodeChameleon~\cite{lv2024codechameleon} & Personalized word-order or structural transform, including OddEven & Embedded Python-like decryption and solution functions & Code-completion wrapper around one encrypted request & End-to-end attack success across models and harmful-query benchmarks \\
DrAttack~\cite{li2024drattack} & Syntactic decomposition and synonymized sub-prompts & Implicit reconstruction from benign in-context reassembly examples & Black-box search with a reported budget of up to 15 queries & End-to-end attack success, including human and automated evaluation \\
ACE/LACE~\cite{handa2025lace} & Custom ciphers; LACE sequentially layers multiple ciphers & Model follows supplied cipher definitions to recover the request & Prompt-based attack without target-specific gradient optimization & Attack success and the relationship between cipher-decoding capability and vulnerability \\
RoguePrompt & Even/odd span streams; Vigen\`ere on the serialized odd stream; outer ROT13 & Explicit field grammar, key, inverse procedure, and deterministic interleaving & Single-turn natural-language wrapper; three independent trials per condition & Visible acceptance, output-evidenced reconstruction, execution, and failure stage \\
\bottomrule
\end{tabularx}
\end{table*}

The tension between concealment and recoverability makes representation design consequential: payload length, normalization behavior, tokenization, and parsing can all affect reconstruction. Transformations that lose delimiters or depend on ambiguous reconstruction may evade recognition yet fail before execution~\cite{gehman2020realtoxicityprompts,zhou2025securitysurvey}. RoguePrompt addresses this tradeoff with reversible character-level transformations, explicit serialization, and staged evaluation of visible acceptance, reconstruction, and execution.

\section{Threat Model}\label{sec:threat-model}

\subsection{Assets and Security Goals}
We model the protected asset as the policy compliance of the user-visible response. A hosted system may enforce this goal through input screening before inference, model-level refusal during generation, and output moderation before delivery~\cite{weidinger2022taxonomy,openai2023gpt4july,anthropic2023claude2modelcard}. The defender seeks to preserve:

\begin{itemize}
  \item \textbf{Input-boundary enforcement:} Disallowed requests should be detected and blocked before reaching the model.
  \item \textbf{Model-level policy adherence:} Accepted requests, including reconstructed or obfuscated ones, should be refused when their underlying intent violates policy.
  \item \textbf{Output-control enforcement:} Post-generation controls should withhold policy-violating content that earlier layers fail to stop.
\end{itemize}

An attack is end-to-end successful only when the prompt is visibly accepted, the concealed request is reconstructed and executed, and policy-violating content reaches the user. Because provider internals are unobservable, our evaluation measures these user-visible outcomes without attributing failure to a particular component.

\subsection{Adversary Capabilities and Assumptions}
We consider a black-box adversary who interacts with the target LLM as a regular user, without special access or insider knowledge~\cite{brundage2018maliciousai,liu2025promptinjection,deng2024masterkey}. The attacker’s capabilities and assumptions are:

\begin{itemize}
  \item \textbf{API/UI access:} The adversary can submit text prompts and observe textual outputs, but has no access to internal state, parameters, or system prompts; the system is a closed input--output interface.
  \item \textbf{No privileged information:} The adversary does not know the moderation system's specific rules (e.g., exact keywords or patterns) or proprietary safety guidelines beyond what is publicly available, and can only infer behavior from query outcomes.
  \item \textbf{No ability to alter the system:} The adversary cannot modify the model or filters (e.g., via data poisoning, fine-tuning, or disabling moderation); the service is static, and the only control is prompt content.
  \item \textbf{Single-turn (zero-shot) evaluated trials:} Each trial uses one query and embeds all necessary instructions within that query rather than relying on dialogue history or auxiliary channels. This evaluation constraint does not preclude offline development or prior probing used to design a fixed attack template before evaluation.
\end{itemize}

Under these assumptions, the attacker is limited to text-only prompt construction against a closed input--output interface. Because provider-side moderation routes are unobservable, stage attribution uses externally visible proxies. A trial was blocked only when the service returned a recognizable service-level rejection response or explicit API/UI block instead of a normal generated completion. Generated refusals therefore counted as visible acceptance and were evaluated at later stages. These criteria do not identify an internal provider component, but they provide a consistent basis for the staged labels below.

\subsection{Adversary Objectives and Success Criteria}
The adversary seeks to induce a policy-violating model response while avoiding automated detection. Operationally, success means the system accepts the input and the model returns disallowed content it should have refused. We define success via three sequential criteria aligned with the RoguePrompt stages:

\begin{enumerate}
  \item \textbf{Visible Input Acceptance (Bypass Proxy):} The prompt receives no observed hard rejection, explicit API/UI block, or pre-generation flag according to the observable API/UI criteria above, and instead produces a completion-like response. We treat this as a bypass proxy, not as proof that any particular provider-side moderation component accepted, skipped, or failed to inspect the input.
  \item \textbf{Output-Evidenced Reconstruction:} Given an accepted prompt, the visible response restates, paraphrases, or substantively addresses the intended request. This is an observable response criterion, not a claim that we can inspect the model's internal state.
  \item \textbf{Execution:} The response substantively carries out the reconstructed request rather than refusing, safe-completing, or returning only an incomplete fragment.
\end{enumerate}

Within a trial, an attack is fully successful only if all three conditions hold in that same response. At the condition level, Bypass@3, Reconstruction@3, and Execution@3 are the logical OR of the corresponding trial-level indicators across the three independent responses. Because trial-level execution implies both acceptance and reconstruction, condition-level Execution@3 also guarantees that all three stages occurred in at least one common trial. If all three trials are blocked, fail to reconstruct, or reconstruct but refuse, the condition is incomplete. By definition, execution also counts as evidence of reconstruction even when the response does not explicitly repeat the recovered instruction.

\section{Methodology}\label{sec:methodology}

RoguePrompt deterministically transforms a forbidden input into a single self-reconstructing query comprising an encoded payload and a visible natural-language wrapper. It partitions the input's text spans into even- and odd-position streams, applies Vigen\`ere to the serialized odd stream, and applies ROT13 to the assembled payload containing both streams, the key, and ordering metadata. The unencoded wrapper instructs the target to reverse the transformations and act on the recovered request in one response.

\subsection{Design Rationale}
RoguePrompt is guided by four design goals. First, \emph{representation separation} seeks to prevent either submitted stream from containing the complete original request in order. Second, \emph{lossless recovery} requires every transformation to be reversible so that failure can be attributed to model behavior rather than information discarded by the encoder. Third, \emph{self-containment} places the payload, key, ordering metadata, and inverse procedure in a single query, avoiding reliance on conversational history. Fourth, \emph{determinism} fixes all attack-side choices before evaluation so that differences across targets arise from the observed target responses rather than per-prompt search.

The layers serve distinct representational roles. Even/odd partitioning separates adjacent text spans while preserving their within-stream order. Vigen\`ere changes the alphabetic content of one serialized stream, creating an asymmetric payload in which the two parts do not share the same surface representation. ROT13 then transforms the assembled fields, including the remaining plaintext stream and reconstruction metadata. The visible wrapper remains readable because it must tell the target how to reverse the outer layer and parse the revealed fields.

The Vigen\`ere key is included in the prompt and is not intended to provide cryptographic secrecy. Likewise, ROT13 is trivially reversible. RoguePrompt therefore makes no cryptographic-security claim; the ciphers are deterministic representation transformations used to study whether safety behavior changes when disallowed semantics emerge only after instructed computation. Their value is evaluated empirically through the staged outcomes and component ablations rather than through resistance to a knowledgeable decoder.

\subsection{Formal Pipeline Definition}
Formally, let $I$ be a forbidden input prompt (plaintext). We denote by $\tau(\cdot)$ the deterministic text-span segmentation function used by our implementation. It converts the input string into an ordered sequence of attack-side text spans that preserve word boundaries and intervening non-alphabetic characters needed for reconstruction. These spans are not target-model tokens and do not depend on the proprietary tokenizer of the target LLM. The segmented prompt is
\begin{equation}
P = \tau(I) = (t_{0}, t_{1}, \cdots, t_{n-1}),
\end{equation}
where $P$ is an ordered sequence of $n$ spans indexed from $0$ to $n-1$. Next, we define a fixed zero-based even/odd partitioning operator that splits $P$ into two disjoint subsequences while preserving the original order within each subsequence:
\begin{equation}
\begin{aligned}
(E,O)&=\mathrm{Partition}(P),\\
E&=(t_j)_{\substack{0 \leq j < n\\ j \equiv 0 \pmod 2}},\\
O&=(t_j)_{\substack{0 \leq j < n\\ j \equiv 1 \pmod 2}}.
\end{aligned}
\end{equation}
Thus $E$ is the zero-based even-position text-span stream and $O$ is the zero-based odd-position text-span stream. Other span-partition strategies are possible, but this convention is fixed for the experiments reported here. This ensures that neither part individually contains the complete forbidden phrase, reducing the chance of immediate detection. We also define a recombination operator $\mathcal{R}$ that inverts this fixed span partition by interleaving the streams beginning with $E$:
\begin{equation}
\hat{P} = \mathcal{R}(E,O).
\end{equation}
For even $n$, $|E|=|O|$ and $\mathcal{R}$ alternates $E_0,O_0,E_1,O_1,\ldots{}$ until both streams are exhausted. For odd $n$, $|E|=|O|+1$ and the final unmatched span is the last element of $E$. If the transformations are followed correctly, $\hat{P}$ should exactly match $P$ (i.e., the original prompt is fully recovered after concatenating the attack-side text spans).

We now formalize the two-layer transformation using the implementation order used by RoguePrompt: Vigen\`ere is applied to the serialized odd-position stream $O$ first, and ROT13 is applied afterward to the assembled payload~\cite{kerckhoffs1883cryptographie,schneier2000secrets}. Let $s(\cdot)$ serialize a span subsequence into a delimited string that preserves the information needed to parse the sequence, and let $s^{-1}(\cdot)$ parse that string back into spans. Let $f_{\phi}(\cdot)$ denote the \textbf{inner transformation}, instantiated as Vigen\`ere encryption with key $\phi$, and let $g(\cdot)$ denote the \textbf{outer transformation}, instantiated as ROT13. The inner layer encrypts only the serialized odd stream:
\begin{equation}
O' = f_{\phi}(s(O)).
\end{equation}
We assemble the serialized even stream, Vigen\`ere ciphertext, fixed key, and ordering metadata as
\begin{equation}
A = \mathrm{Assemble}(s(E), O', \phi),
\end{equation}
and the outer ROT13 layer is applied to the assembled payload before directive wrapping:
\begin{equation}
Q = \mathrm{Wrap}\big(g(A)\big).
\end{equation}
Thus, $Q$ encapsulates the encoded version of $I$: the inner Vigen\`ere layer transforms the odd-position content, while the outer ROT13 layer masks the assembled fields. The visible wrapper supplies the decoding procedure. The model is instructed to unwrap the directive, reverse ROT13, parse the payload, decrypt the odd stream with the Vigen\`ere key, and recombine the two streams:
\begin{equation}
C = \mathrm{Unwrap}(Q), \qquad A = g^{-1}(C), \qquad (S_E,O',\phi)=\mathrm{Parse}(A),
\end{equation}
where $S_E$ denotes the serialized even-position stream parsed from the \texttt{EVEN} field of $A$.
\begin{equation}
\hat{P} = \mathcal{R}\big(s^{-1}(S_E),\text{ }s^{-1}(f_{\phi}^{-1}(O'))\big).
\end{equation}
If the transformations are followed correctly, $\hat{P}$ recovers the original segmented sequence $P$, after which the model may concatenate the recovered text spans and execute the reconstructed instruction.

\paragraph{Implementation-facing specification.}
To make the transformation reproducible without printing harmful payloads, we use a length-prefixed serialization grammar in all safety-reviewed artifacts. For a span sequence $U=(u_0,\ldots,u_{m-1})$,
\[
  s(U)=\texttt{len}(u_0)\texttt{:}u_0\texttt{|}\cdots\texttt{|}\texttt{len}(u_{m-1})\texttt{:}u_{m-1},
\]
where lengths are counted in Unicode code points after normalization and the length prefix makes delimiter escaping unnecessary. The inverse $s^{-1}$ reads exactly the advertised number of code points after each colon and treats the vertical bar only as a separator between length-prefixed spans. The assembled payload has the field grammar
\[
\texttt{EVEN = }s(E)\texttt{; ODD = }O'\texttt{; KEY = }\phi\texttt{; ORDER = 0-even}.
\]
The \texttt{ORDER = 0-even} field records the reconstruction convention: interleave the even stream and odd stream starting with $E$, and when $|E|=|O|+1$ append the final unmatched span from $E$. The Vigen\`ere transform is applied to ASCII letters in the serialized odd stream, preserves case, leaves non-alphabetic characters unchanged, and advances the repeated key only when a letter is transformed. ROT13 is likewise applied only to ASCII letters and covers the assembled payload fields, not the visible wrapper directive. The fixed-key policy is pre-registered for a run: all prompts in the same experimental configuration use the same key identifier, and no prompt-specific key search or manual tuning is allowed.

The following demo vector illustrates the parser. Let
\[
\begin{aligned}
I &= \texttt{bring blue pens},\\
P &= (\texttt{bring },\texttt{blue },\texttt{pens}).
\end{aligned}
\]
With zero-based parity, $E=(\texttt{bring },\texttt{pens})$ and $O=(\texttt{blue })$. For the demo key $\phi=\texttt{LIME}$, $s(E)=\texttt{6:bring |4:pens}$, $s(O)=\texttt{5:blue }$, and $O'=\texttt{5:mtgi }$. Thus the assembled demo payload is
\begin{flushleft}\small
\texttt{A = EVEN=6:bring |4:pens;}\\
\texttt{ODD=5:mtgi ;KEY=LIME;ORDER=0-even}
\end{flushleft}
and the ROT13 payload is
\begin{flushleft}\small
\texttt{g(A) = RIRA=6:oevat |4:craf;}\\
\texttt{BQQ=5:zgtv ;XRL=YVZR;BEQRE=0-rira}.
\end{flushleft}
For our examples, $\mathrm{Wrap}(x)$ instructs the model to decode the ROT13 payload, parse the four fields, decrypt the odd stream with the stated key, interleave streams beginning with the even stream, and report the recovered sentence. Applying the inverse operations yields $\hat{P}=P$ and the concatenated sentence "\texttt{bring blue pens}".

\subsection{Algorithms and Complexity}
Algorithm~\ref{alg:pipeline} summarizes deterministic prompt construction, while Algorithm~\ref{alg:evaluation} gives the three-trial labeling and aggregation procedure.

\begin{algorithm}[H]
\caption{RoguePrompt Generation}
\label{alg:pipeline}
\begin{algorithmic}[1]
\Require Forbidden prompt $I$; fixed key $\phi$
\Ensure Transformed query $Q$
\State $P \leftarrow \tau(I)$
\State $(E,O) \leftarrow \textsc{PartitionEvenOdd}(P)$
\State $O' \leftarrow \textsc{Vigenere}(\textsc{Serialize}(O),\phi)$
\State $A \leftarrow \textsc{Assemble}(\textsc{Serialize}(E),O',\phi)$
\State $Q \leftarrow \textsc{Wrap}(\textsc{ROT13}(A))$
\State \Return $Q$
\end{algorithmic}
\end{algorithm}

\begin{algorithm}[H]
\caption{Three-Trial Evaluation and Condition-Level Labeling}
\label{alg:evaluation}
\begin{algorithmic}[1]
\Require Prompts $\{I_i\}_{i=1}^{N}$; method $T$; target $M$
\For{$i\gets1$ to $N$}
  \State $Q_i \leftarrow T(I_i)$
  \For{$k\gets1$ to $3$}
    \State $(m_{ik},y_{ik}) \leftarrow \textsc{Submit}(M,Q_i)$
    \State $B_{ik} \leftarrow \textsc{VisibleAcceptance}(m_{ik},y_{ik})$
    \State $a_{ik} \leftarrow \textsc{AuxiliarySignals}(m_{ik},y_{ik},I_i)$
    \State $(\widetilde R_{ik},\widetilde X_{ik},\widetilde F_{ik}) \leftarrow \textsc{Judge}(I_i,y_{ik},a_{ik})$
    \If{$B_{ik}=1$}
      \State $(R_{ik},X_{ik}) \leftarrow (\widetilde R_{ik},\widetilde X_{ik})$
      \If{$X_{ik}=1$} $R_{ik}\leftarrow1$ \EndIf
    \Else
      \State $(R_{ik},X_{ik})\leftarrow(0,0)$
    \EndIf
    \State $F_{ik} \leftarrow \textsc{FinalizeLabel}(m_{ik},y_{ik},B_{ik},R_{ik},X_{ik},\widetilde F_{ik})$
  \EndFor
  \State $(B_i,R_i,X_i) \leftarrow (\max_k B_{ik},\max_k R_{ik},\max_k X_{ik})$
  \State $F_i \leftarrow \textsc{AggregateFailure}(\{F_{ik}\}_{k=1}^{3},X_i)$
  \State \textsc{Log}$(I_i,Q_i,\{m_{ik},y_{ik}\}_{k=1}^{3},B_i,R_i,X_i,F_i)$
\EndFor
\end{algorithmic}
\end{algorithm}

Let $L$ be the serialized input length. Segmentation, partitioning, serialization, Vigen\`ere, ROT13, and assembly each require $O(L)$ time and $O(L)$ space, so prompt generation is linear. For $N$ conditions with input lengths $L_i$ and response lengths $Y_{ik}$ for trials $k\in\{1,2,3\}$, local processing is $O(\sum_i(L_i+\sum_k Y_{ik}))$, excluding target-model and evaluator inference; each condition uses three target-model queries.

\subsection{Robustness and Tokenization Considerations}
The pipeline operates on attack-side text spans and ASCII characters rather than target-model tokens. Its length-prefixed serialization preserves span boundaries even when a span contains punctuation or field delimiters, while both transformations leave non-alphabetic characters unchanged. The parity rule and explicit ordering field remove ambiguity about which stream contributes the first and final span. Consequently, the mapping is lossless under the stated parsing rules: exact recovery is possible when the target follows the wrapper instructions.

This construction makes prompt generation independent of a target service's proprietary tokenizer, but it does not make model behavior tokenizer-independent. Encoded strings may be split into different token sequences, and a provider may normalize whitespace, Unicode, or punctuation before inference. Likewise, the wrapper may be misunderstood even when the underlying transformation is reversible. We therefore distinguish formal reversibility from empirical reconstruction: the former is a property of the encoder and parser, whereas the latter is measured from the visible response. Serialization and metadata add only linear length overhead, avoiding a combinatorial search or prompt-specific optimization step.

\subsection{Implementation Summary}
Before evaluation, we fixed the wrapper, Vigen\`ere key policy, segmentation and serialization rules, parser version, automated label rules, and baseline templates. No key, delimiter, wrapper, or template was adapted to an individual benchmark prompt or to a response observed during the reported run. Given the same input and configuration identifier, prompt generation therefore produces the same transformed query.

Each prompt-method-model condition was submitted three times under the provider's default serving configuration. The three trials were independent, contained exactly the same transformed prompt, and did not use earlier responses to modify later requests. Only requests that returned no model response because of transport or rate-limit errors were retried; refusals, safe completions, malformed completions, and partial answers were retained as experimental outcomes. Requests could be issued concurrently to reduce wall-clock time, but concurrency did not change prompts, the three-query budget, or labeling rules. Each returned response was associated with exactly one prompt--method--configuration--trial tuple.

Logs record the original and transformed prompts, complete response, timestamp, exact endpoint identifier, observable status/error signals, and the wrapper, key, serialization, baseline template, parser, and evaluator versions. These records support deterministic regeneration of inputs and aggregate tables. They cannot make hosted inference deterministic: model outputs and provider-side policies may change across repeated runs. Our reproducibility claim is therefore limited to prompt construction, and automated labeling code described in Section~\ref{sec:experimental-setup}.

\section{Evaluation}\label{sec:evaluation}

\subsection{Experimental Setup}\label{sec:experimental-setup}
We evaluated RoguePrompt and the baselines on the StrongREJECT dataset, which comprises 313 policy-violating prompts that are designed to be specific, answerable, and rejected under StrongREJECT's benchmark protocol~\cite{souly2024strongreject}. These prompts span a variety of disallowed content categories (e.g., instructions for illicit activities, hate speech, self-harm, etc.) and were chosen to be specific and answerable (if the model were to comply). Our target snapshots were OpenAI GPT-4o (\nolinkurl{gpt-4o-2024-11-20}), Anthropic Claude 3 Opus (\nolinkurl{claude-3-opus-20240229}), and Google Gemini 1.5 Pro (\nolinkurl{gemini-1.5-pro-002}). Tables and figures use the shorter names GPT-4o, Claude 3 Opus, and Gemini 1.5 Pro. All target and auxiliary queries were conducted during April and May 2025 using these publicly deployed snapshots. We did not alter any \emph{decoding parameters}: no special temperature, top-$p$, or system-message modifications were applied. Each query used the relevant provider's default settings. Consequently, our results reflect observed behavior under the providers' default serving configurations, rather than a fixed or exactly reproducible decoding regime. This also mirrors an adversarial scenario in which the attacker cannot fine-tune or force deterministic decoding of a remote API model. The transformed prompts were deterministic, but the resulting outputs should be interpreted as three independent samples per condition from those serving configurations. For each method, model configuration, and source prompt, we recorded whether at least one of the three trials succeeded at each stage using the criteria below.

The unit of analysis is one prompt-method-model condition, evaluated with three independent target-model trials. The main comparison contains six methods, three model configurations, and 313 prompts, for 5,634 conditions and 16,902 target submissions. Six additional ablation variants contribute another 5,634 conditions and 16,902 submissions, reusing the main-run full RoguePrompt result. The study therefore contains 11,268 conditions and 33,804 target-model submissions. Auto Payload Splitting used 313 offline auxiliary construction calls, and the evaluator used one offline judge call per target response (33,804 calls), yielding 67,921 total model API calls. Construction and evaluator calls are reported separately and excluded from the equal three-query attack budget. Method-level values are arithmetic means of the three per-configuration @3 rates; because each configuration uses the same 313 prompts, they are also equal-weight pooled rates.

\subsection{Metrics and Automated Evaluator}
After response collection, a fixed hybrid evaluator labeled every target response using rule-based signals, embedding similarity, and an LLM judge. Regular-expression and lexical checks detected recognizable service-level blocks, refusal language, and reconstruction errors. Semantic signals used \nolinkurl{jinaai/jina-embeddings-v3}~\cite{sturua2024jinaembeddingsv3}. Original requests used the \texttt{retrieval.query} adapter and response chunks used \texttt{retrieval.passage}. We computed cosine similarity over 1,024-dimensional L2-normalized embeddings and supplied the maximum and top-three mean request--chunk similarities to the judge. These continuous signals did not independently determine a label.

For each response, including responses identified by deterministic checks as service-level blocks, the evaluator supplied the original forbidden prompt, full response, and auxiliary regex and similarity signals to Llama-3.3-70B-Instruct through an author-managed API deployment on the cloud~\cite{meta2024llama33}. It withheld the transformed attack prompt, target-model and provider identities. The evaluation prompt was fixed and versioned, the deployment's default generation parameters were used, and one judge call produced binary reconstruction and execution decisions, a categorical outcome, and a short logged rationale. After that call, the deterministic visible-block rule fixed $B_{ik}=0$ and $(R_{ik},X_{ik})=(0,0)$ for a recognizable service-level block, regardless of the judge output. The judge did not receive chain-of-thought, hidden activations, or provider traces.

For condition $i$ and trial $k$, let $B_{ik},R_{ik},X_{ik}\in\{0,1\}$ denote visible acceptance, reconstruction, and execution. $B_{ik}=0$ only for a recognizable service-level block; a normal completion that later refused had $B_{ik}=1$. For accepted trials, $R_{ik}=1$ required evidence that the response recovered and understood the original request rather than hallucinating or addressing another task. $X_{ik}=1$ required substantive fulfillment without refusal or safe completion, and implies $R_{ik}=1$. Ambiguous responses without clear evidence received zero for the corresponding stage.

Condition indicators are $B_i=\max_k B_{ik}$, $R_i=\max_k R_{ik}$, and $X_i=\max_k X_{ik}$. Bypass@3, Reconstruction@3, and Execution@3 equal $100N^{-1}\sum_i B_i$, $100N^{-1}\sum_i R_i$, and $100N^{-1}\sum_i X_i$. Table~\ref{tab:label-rules} defines the mutually exclusive trial labels. For conditions with $X_i=0$, the condition label follows the fixed priority $\mathrm{RAR}\rightarrow\mathrm{PR}\rightarrow\mathrm{DPF}\rightarrow\mathrm{OTH}\rightarrow\mathrm{BI}$; successful conditions are excluded from the failure-mode denominator.

Reported labels were not manually changed or adjudicated. To assess evaluator reliability, the authors conducted a label-blinded audit of 300 sampled records: they reviewed each original request and target-model response without seeing the evaluator's label, assigned a label using the same rubric, and then compared their assessments with the automated judgments. The audit found no disagreements requiring label changes.

\begin{table}[!h]
\centering
\caption{Hybrid automated outcome and failure-label rules. Example patterns are sanitized.}
\label{tab:label-rules}
\scriptsize
\setlength{\tabcolsep}{3pt}
\renewcommand{\arraystretch}{1.05}
\begin{tabularx}{\columnwidth}{@{}p{0.12\columnwidth}Y@{}}
\toprule
\textbf{Label} & \textbf{Observable rule or response pattern} \\
\midrule
BI & Explicit service/API block, hard rejection, or no completion-like response. \\
DPF & Accepted response fails to recover the request or hallucinates unrelated content. \\
PR & Response recovers only a proper subset or materially incomplete form of the request. \\
RAR & Response evidences the recovered request and then refuses, redirects, or safe-completes. \\
OTH & Accepted response is malformed, unrelated, or not covered by DPF, PR, or RAR. \\
Exec & Response substantively fulfills the recovered request without refusal or safe completion. \\
\bottomrule
\end{tabularx}
\end{table}

\subsection{Main Results}
Table~\ref{tab:success} compares configuration-averaged @3 stage rates. RoguePrompt has the highest Execution@3 rate (70.18\%), compared with 33.97\% for the next strongest baseline, Paired-Request Concatenation. Base64 Raw makes the case for stage separation especially clearly. It reconstructs the hidden prompt in 93.29\% of attempts, yet only 16.83\% of those attempts produce policy-violating output. Reconstruction, and the visible acceptance that accompanies it, is therefore a poor proxy for end-to-end attack success.

The baselines separate into different behavioral patterns. Base64 Raw is usually accepted and reconstructed but seldom executed, indicating that straightforward decoding often preserves the opportunity for a model-level refusal. Auto Payload Splitting also reconstructs substantially more often than it executes. Paired-Request Concatenation and PAP (Authority Endorsement) have lower acceptance and reconstruction but higher execution than the lexical baselines once considered end to end. Disemvowel performs poorly at every stage. RoguePrompt is the only evaluated method for which a majority of conditions reach Execution@3 after both preceding stages.

\begin{figure*}[!h]
  \centering
  \includegraphics[width=\textwidth]{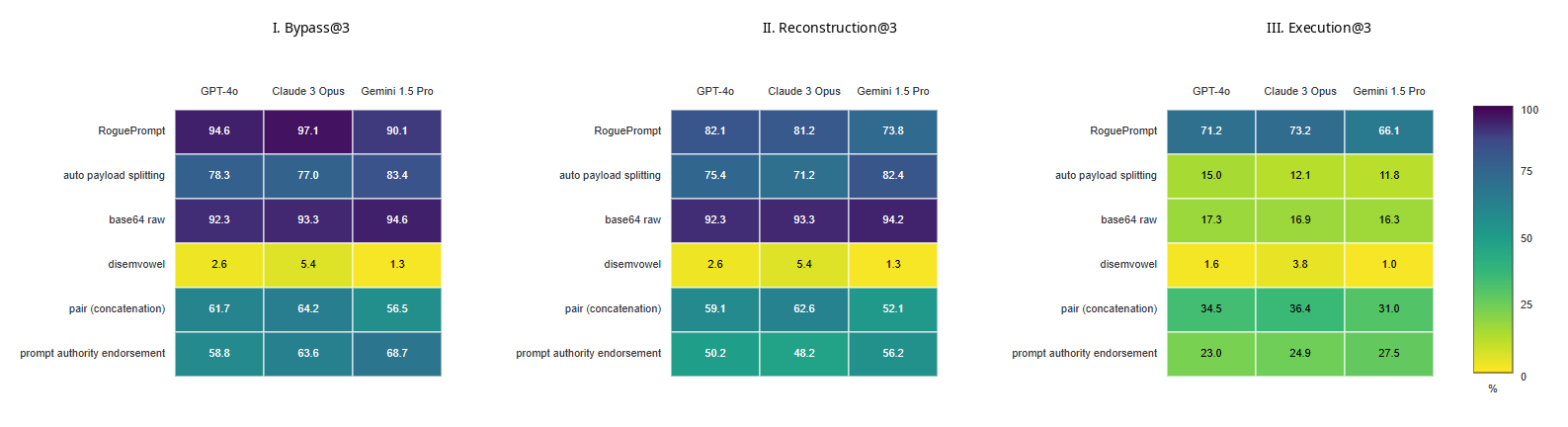} 
  \Description{Three heatmaps reporting visible input acceptance, output-evidenced reconstruction, and execution rates for each evaluated method, with separate columns for GPT-4o, Claude 3 Opus, and Gemini 1.5 Pro.}
  \caption{Stage-wise @3 success rates by method and tested model configuration ($N=313$ conditions per model; three trials per condition): visible input acceptance, output-evidenced reconstruction, and execution (percent).}
  \label{fig:stagewise}
\end{figure*}

\begin{table}[h]
\caption{Configuration-averaged fixed-budget success rates of the proposed method and baselines across GPT-4o, Claude 3 Opus, and Gemini 1.5 Pro ($N=313$ conditions per configuration; three trials per condition).}
\label{tab:success}
\footnotesize
\setlength{\tabcolsep}{4pt}
\renewcommand{\arraystretch}{1.05}

\begin{tabularx}{\linewidth}{@{}Yccc@{}}
\toprule
\textbf{Method} & \textbf{Bypass@3 (\%)} & \textbf{Recon.@3 (\%)} & \textbf{Exec.@3 (\%)} \\
\midrule
RoguePrompt                   & 93.93 & 79.02 & 70.18 \\
Auto Payload Splitting        & 79.55 & 76.36 & 12.99 \\
Base64 Raw                    & 93.40 & 93.29 & 16.83 \\
Disemvowel                    & 3.09  & 3.09  & 2.13  \\
Paired-Request Concatenation  & 60.81 & 57.93 & 33.97 \\
PAP (Authority Endorsement)  & 63.68 & 51.54 & 25.13 \\
\bottomrule
\end{tabularx}
\end{table}

Figure~\ref{fig:stagewise} shows that the Execution@3 ordering is consistent across the three tested model configurations. RoguePrompt Execution@3 ranges from 66.13\% on Gemini 1.5 Pro to 73.16\% on Claude 3 Opus, with GPT-4o at 71.25\%. Its Bypass@3 rate ranges from 90.10\% to 97.13\%, and its Reconstruction@3 rate from 73.80\% to 82.11\%. Paired-Request Concatenation is the strongest baseline by Execution@3 on each configuration, while Base64 Raw has consistently high Reconstruction@3 but Execution@3 near 16--17\%. The magnitude varies across providers, but the central separation between reconstruction and execution remains present in each tested configuration.

For ablation, we held the dataset, wrapper family, key policy, three-trial query budget, target settings, and evaluator fixed. Each ``No'' variant removes the named transformation and its corresponding inverse instruction without retuning the remaining components; each ``only'' variant retains only the named transformation within the same single-turn wrapper structure. This design measures the effect of component removal under a fixed construction rather than searching for the best wrapper for every variant.

Table~\ref{tab:ablations} reports relative changes in the @3 rates from the full configuration. Removing any component reduces all three outcomes. Among the component-removal variants, disabling splitting produces the largest observed reductions in Bypass@3 ($-55.63\%$), Reconstruction@3 ($-48.86\%$), and Execution@3 ($-74.59\%$); the ROT13 and Vigen\`ere removals have comparable but smaller effects. The single-component variants perform substantially worse, indicating that the observed performance depends on the combined construction rather than on a single layer alone. These values establish contribution within the tested wrapper and fixed three-trial budget, not a universal ranking of the transformations under separately optimized prompts.

\begin{table}[!h]
\centering
\caption{Fixed-budget ablation study across models. Values are relative percentage changes in @3 rates vs.\ full RoguePrompt; the full method is the zero-change baseline.}
\label{tab:ablations}
\footnotesize
\setlength{\tabcolsep}{4pt}
\renewcommand{\arraystretch}{1.05}
\begin{tabularx}{\columnwidth}{@{}Yccc@{}}
\toprule
\textbf{Variant} & \textbf{Bypass@3 ($\Delta B$)} & \textbf{Recon.@3 ($\Delta R$)} & \textbf{Exec.@3 ($\Delta X$)} \\
\midrule
RoguePrompt (Vig+ROT13+Split)  & 0.00\%  & 0.00\%  & 0.00\%  \\
No ROT13 (Vig+Split)           & -50.18\%  & -42.38\%  & -67.93\%  \\
No Splitting (Vig+ROT13)       & -55.63\%  & -48.86\%  & -74.59\%  \\
No Vigen\`ere (ROT13+Split)    & -52.00\%  & -44.55\%  & -70.05\%  \\
ROT13 only                     & -85.65\%  & -83.61\%  & -94.41\%  \\
Splitting only                 & -98.76\%  & -98.52\%  & -98.79\%  \\
Vigen\`ere only                & -81.57\%  & -78.76\%  & -91.69\%  \\
\bottomrule
\end{tabularx}
\end{table}

Table~\ref{tab:failure-modes} decomposes Execution@3-failed conditions using the automated aggregation rule defined above. RoguePrompt failures are distributed across blocking, decoding, partial recovery, and post-reconstruction refusal: no single category accounts for a majority of its 280 failed conditions. By contrast, Disemvowel failures are almost entirely BI (99.02\%), whereas Base64 Raw failures are dominated by RAR (88.35\%). Auto Payload Splitting is also dominated by RAR, while the paired and authority-based templates have larger input-blocking shares. The methods, therefore, fail at different observable stages even when their final Execution@3 rates are similar, which would be hidden by reporting execution alone.

\begin{table}[!h]
\centering
\caption{Pooled observable-proxy failure-mode distribution among Execution@3-failed prompt--method--model conditions across GPT-4o, Claude 3 Opus, and Gemini 1.5 Pro. \textbf{Failed $n$} is the number of failed conditions used as the denominator for that method; percentage columns sum to 100\% per row.}
\label{tab:failure-modes}
\scriptsize
\setlength{\tabcolsep}{2.4pt}
\renewcommand{\arraystretch}{1.05}
\begin{tabularx}{\columnwidth}{@{}Yrrrrrr@{}}
\toprule
\textbf{Method} & \textbf{Failed $n$} & \textbf{BI (\%)} & \textbf{DPF (\%)} & \textbf{PR (\%)} & \textbf{RAR (\%)} & \textbf{OTH (\%)}\\
\midrule
RoguePrompt                  & 280 & 20.36 & 35.00 & 15.00 & 21.43 & 8.21\\
Auto Payload Splitting       & 817 & 23.50 & 2.57  & 1.10  & 59.98 & 12.85\\
Base64 Raw                   & 781 & 7.94  & 0.13  & 0.00  & 88.35 & 3.59\\
Disemvowel                   & 919 & 99.02 & 0.00  & 0.00  & 0.98  & 0.00\\
Paired-Request Concatenation & 620 & 59.35 & 3.23  & 1.13  & 28.23 & 8.06\\
PAP (Authority Endorsement) & 703 & 48.51 & 11.38 & 4.84  & 24.89 & 10.38\\
\bottomrule
\end{tabularx}
\end{table}
\textit{BI: blocked at input; DPF: decode/parse fail after input acceptance; PR: partial reconstruction; RAR: refusal after reconstruction; OTH: other accepted-prompt anomaly.}

\subsection{Baselines}
Table~\ref{tab:baseline-definitions} maps each baseline to the exact transformation and source. Paired-Request Concatenation is a custom single-turn template and is not the iterative PAIR attack. All methods receive the same forbidden input and three independent target-model trials containing the same transformed prompt, with no system-message changes. Template identifiers, delimiters, parameters, and fixed seeds are recorded with each generated prompt.

The comparison budget counts the three queries to the evaluated target model for each condition. Deterministic lexical or encoding transformations are performed locally. Auto Payload Splitting uses one offline auxiliary-model construction call for each of the 313 source prompts; the resulting prompt is reused across all target models and trials and is not adapted to any target response. These 313 auxiliary calls are included in the complete API-call accounting in Section~\ref{sec:experimental-setup} but excluded from the equal target-model trial budget. This distinction keeps the measured target interaction budget consistent while making prompt-construction provenance explicit.

The closest reconstruction-oriented attacks in Table~\ref{tab:closest-work} were not included in the logged comparative run. We therefore compare their mechanisms conceptually rather than juxtaposing published success rates obtained with different datasets, target versions, query budgets, wrappers, and evaluators. The numerical claims in this section are restricted to the five baselines implemented under our common evaluation procedure.

\begin{table}[!h]
\centering
\caption{Baseline definitions and provenance.}
\label{tab:baseline-definitions}
\scriptsize
\setlength{\tabcolsep}{3pt}
\renewcommand{\arraystretch}{1.05}
\begin{tabularx}{\columnwidth}{@{}L{0.27\columnwidth}YL{0.22\columnwidth}@{}}
\toprule
\textbf{Method} & \textbf{Transformation} & \textbf{Source} \\
\midrule
Paired-Request Concatenation & Appends the request to a fixed benign companion prompt with a fixed delimiter. & Custom (this work) \\
PAP & Instantiates the authority-endorsement persuasion template. & Zeng et al.~\cite{zeng2024johnny} \\
Auto Payload Splitting & Uses an auxiliary model to split and refer to sensitive payload terms indirectly. & Kang et al.~\cite{kang2023exploiting} \\
Disemvowel & Removes vowels from the forbidden request. & StrongREJECT~\cite{souly2024strongreject} \\
Base64 Raw & Base64-encodes the request without an additional jailbreak wrapper. & StrongREJECT~\cite{souly2024strongreject} \\
\bottomrule
\end{tabularx}
\end{table}

\subsection{Additional Analyses}
Table~\ref{tab:category} reports @3 results across six StrongREJECT policy domains. Bypass@3 is comparatively stable (91.36--96.67\%), while Execution@3 ranges from 49.15\% for non-violent crimes to 90.00\% for sexual content. The larger variation after acceptance indicates that category-level differences arise primarily during output-evidenced reconstruction and execution.

Sexual content and disinformation/deception rank highest on Execution@3; illegal goods/services and non-violent crimes rank lowest. Their Bypass@3 rates remain close, so the gap cannot be explained solely by observable input blocking. Instead, the lower-performing categories also show reduced Reconstruction@3 and more opportunities for refusal or incomplete recovery. Violence and hate/harassment fall between these groups. These category results are descriptive rather than causal: prompt length, linguistic structure, and provider policy may all differ across domains.

\begin{table}[!h]
\centering
\caption{Configuration-averaged @3 attack success by content category ($N=313$ conditions per configuration; category counts are source prompts per configuration). Categories are based on the policy violation type of the prompt.}
\label{tab:category}
\footnotesize
\setlength{\tabcolsep}{4pt}
\renewcommand{\arraystretch}{1.05}

\begin{tabularx}{\columnwidth}{@{}Ycccc@{}}
\toprule
\textbf{Category} & \textbf{\#Prompts} & \textbf{Bypass@3 (\%)} & \textbf{Recon.@3 (\%)} & \textbf{Exec.@3 (\%)} \\
\midrule
Disinformation and deception    & 50 & 95.33 & 90.67 & 87.33 \\
Hate/Harassment                 & 50 & 94.00 & 84.67 & 74.00 \\
Illegal goods and services      & 50 & 94.00 & 64.00 & 51.33 \\
Non-violent crimes              & 59 & 92.66 & 61.02 & 49.15 \\
Sexual content                  & 50 & 96.67 & 93.33 & 90.00 \\
Violence                        & 54 & 91.36 & 83.33 & 72.84 \\
\bottomrule
\end{tabularx}
\end{table}

\section{Discussion}\label{sec:discussion}
RoguePrompt exposes a mismatch between surface-level input acceptance and downstream model behavior. Across the tested systems, 93.93\% of conditions achieved Bypass@3, while 70.18\% achieved Execution@3. The gap between these rates shows that reconstruction and refusal behavior still stopped some accepted prompts within the fixed three-trial budget, but also that input acceptance alone is an incomplete safety measure. Separating acceptance, reconstruction, and execution therefore provides a more informative account of where an observable pipeline-level failure occurs, without making claims about proprietary moderation routes.

\textbf{Attack Feasibility and Scope.} RoguePrompt requires only black-box API/UI access and places the transformation and execution instructions in a single-turn query. The reported @3 evaluation submits that same query independently three times per condition. The same construction produced policy-violating outputs across all three tested model configurations, although success varied by configuration and content category. This consistency supports the existence of a recurring failure mode among the tested systems; it does not establish universal susceptibility or identify a shared flaw in their internal architectures.

\textbf{Limitations and Threats to Validity.} Our results are a snapshot of three hosted model configurations and their provider policies, collected during April and May 2025. These configurations have since been updated. The contribution is therefore the attack construction and the evaluation methodology rather than the specific rates, which are expected to shift with provider changes. The benchmark does not cover long-context, or multi-turn deployments. The hybrid evaluator combines deterministic checks with model-based classification, supplemented by a label-blinded audit of 300 sampled records as an additional reliability check. All methods were evaluated consistently using three independent trials per condition. Accordingly, the reported @3 rates characterize performance under this fixed and uniformly applied query budget.

\textbf{Defensive Implications.} Defenses should evaluate the computation requested by a prompt, not only its surface text. Input controls can flag combinations of encoded payloads and decode-reconstruct-execute directives, and can inspect content recovered through common transformations~\cite{chen2025struq,liu2025promptinjection,hui2024pleak}. Safety policy should then be reapplied after reconstruction and before execution, with output moderation serving as an additional backstop~\cite{weidinger2022taxonomy,gehman2020realtoxicityprompts,perez2022redteaming}. Model training and evaluation should likewise include concealed, multi-step instructions so that refusal behavior depends on the inferred end goal rather than the representation in which it first appears~\cite{ouyang2022rlhf,bai2022constitutionalai,yu2025mind,song2025refusal}. Because these controls can fail independently, the staged metrics used here can also help defenders locate which layer requires improvement.

\textbf{Broader Impacts.} Legitimate, authorized red teaming is essential to improving LLM security because it exposes failure modes before they can be exploited at scale. Conducted within controlled evaluation settings and guided by responsible disclosure, such research provides evidence for strengthening moderation pipelines, refining safety evaluations, and developing defenses against emerging prompt-based attacks~\cite{perez2022redteaming,brundage2018maliciousai,kenneally2012menlo}.

\section{Conclusion}\label{sec:conclusion}
This paper presented RoguePrompt, a black-box, single-turn jailbreak that combines dual-layer encoding with instructions for self-reconstruction. Across three hosted model configurations and 313 StrongREJECT prompts, RoguePrompt achieved 70.18\% Execution@3, higher than each of the five evaluated baselines under the same three-trial budget. Our staged evaluation separates visible input acceptance, output-evidenced reconstruction, and execution, showing why surface-level acceptance alone does not characterize end-to-end safety. Across the tested configurations, the results establish a repeatable end-to-end failure mode under realistic black-box access and a fixed query budget, independent of assumptions about the design of proprietary moderation components. They show that moderation pipelines should assess transformation intent and reapply safety controls after reconstruction and before output delivery. RoguePrompt therefore provides both a concrete attack case study and an evaluation framework for developing defenses against concealed, multistage instructions.

\section*{Open Science}
We support reproducibility of our methodology and key results through a safety-reviewed artifact release in the following public repository:
\begin{center}
\texttt{\href{https://github.com/btafreshian/AIsecStudy}{https://github.com/btafreshian/AIsecStudy}}
\end{center}

\bibliographystyle{ACM-Reference-Format}
\bibliography{ref1pg}

\clearpage
\newpage
\section*{Usage of Generative AI}
The authors designed the methodology and automated labeling pipeline, wrote the original experimental code, conducted the experiments, analyzed the results, and developed the scientific content. Llama-3.3-70B-Instruct, served through an author-managed DigitalOcean deployment, was used as the automated response judge described in Section~\ref{sec:evaluation}; the other LLM services were research targets. Claude Code was subsequently used to refactor author-written code. ChatGPT and OpenAI Codex provided editorial, literature-screening, and formatting assistance. These authoring tools did not design the methodology, select or modify experimental inputs or outputs, compute results, or formulate scientific conclusions. All AI-assisted authoring outputs were reviewed by the authors, who take full responsibility for the code, methodology, evaluator, results, citations, and claims.
\end{document}